\documentstyle[prl,preprint,aps]{revtex}

\begin{document}

\title {Effects of a weak disorder on two coupled Hubbard chains}

\author{E. Orignac and T. Giamarchi}
\address{Laboratoire de Physique des Solides, Universit{\'e} Paris--Sud,
                   B{\^a}t. 510, 91405 Orsay, France\cite{junk}}
                   \date{\today}
\maketitle

\begin{abstract}
We consider the effect of weak nonmagnetic disorder
on two chains of interacting fermions (with and without spins)
coupled by interchain hopping.
For the spinless case, interchain hopping increases localization
for repulsive interactions but {\it stabilizes} the s-wave
superconducting phase for attractive interactions.
For the case with spin, the d-wave phase arising from purely repulsive
interactions in the clean system is destroyed by an infinitesimal disorder
while
for attractive interactions, the s-wave superconductivity is more
resistant to disorder than in the one-chain case.
\end{abstract}
\pacs{to be added}

%\begin{multicols}{2}
\narrowtext

One dimensional electronic systems are  known to be the simplest
realizations of non-fermi liquids, and to have generic properties known
as Luttinger liquids
\cite{emery_revue_1d,solyom_revue_1d,haldane_bosonisation}.
Despite the good understanding of purely one
dimensional systems, the effects of interchain hopping, allowing to go
from one to higher (two or three) dimensions are much less known.
Whether such hopping is relevant and drives the system towards a Fermi
liquid fixed point or whether non-fermi liquid properties can be
retained even in presence of finite hopping is still a controversial
issue
\cite{anderson_luttinger,bourbon_couplage,schulz_trieste,yakovenko}.
A theoretical understanding of quasi one-dimensional
strongly correlated electronic systems (especially the crossover from
Luttinger to Fermi liquid) would
be relevant for the physics of organic conductors and may perhaps give
some insights for High-Tc superconductors.
Consequently, there has been in the recent years,
a growing interest in systems of coupled
interacting electron chains, and in particular in systems of two chains.
They present the advantage to allow a careful study of the
effects of the hopping, as well as to be tractable by powerful
analytical
\cite{fabrizio_2ch_rg,finkelstein_2ch,schulz_2chains,balents_2ch,nagaosa_2ch,nersesyan_2ch}
and numerical techniques
\cite{noack_dmrg_2ch,dagotto_lanczos_2ch,poilblanc_2ch,tsunegutsu_2ch}.
In addition there exists good
experimental realizations of such systems. For
example
$\rm Sr_{n-1}Cu_{n+1}O_{2n}$ \cite{takano_srcuo} and $\rm VO_2P_2O_7$
\cite{eccleston_vo2p2o7} compounds are very good
realizations of coupled spin chains.
Upon doping, such compounds will give coupled Hubbard chains.
Although the complete phase diagram of such systems is still under
study, a generic property of two coupled chains system
is the appearance of a
$d$-wave like superconducting phase for repulsive interactions.

In this work we study the effects of non-magnetic disorder on such
two chains systems, both for the case of spinless electrons and for
electrons with spins. Such a study has a double interest: in
experimental systems, disorder will be present, and it is therefore
essential to know the stability of the phases found in
the pure system. It is now well known that for a strictly one dimensional
system, disorder has extremely strong effects and an arbitrarily weak
disorder destroys superconductivity except for exceedingly attractive
interactions \cite{giamarchi_loc}. In addition, on a more theoretical
level, the two chain
problem is the simplest one to study the effects of interchain hopping
onto the Anderson localization in presence of interactions, giving some
clues to the unsatisfactorily
understood physics of such transition in more than one dimension.
We show here that for the spinless model the superconducting phase for
attractive interactions is {\em stable} towards weak disorder at the
opposite of what happens for a purely one dimensional system. For the
model with spins and attractive interactions an arbitrarily weak disorder
destroys the superconductivity  if the interactions are not attractive
enough as in the one chain case. Nevertheless, the threshold in interaction
strength to induce superconductivity  is much smaller for two chains.
In particular it can now be reached for a pure Hubbard
attraction at variance to the one chain case \cite{giamarchi_persistent_1d}.
On the other hand the
$d$-wave type superconductivity found for repulsive interactions is
completely unstable with respect to arbitrarily weak disorder.
In two spinless chains, attractive interactions {\it reduce}
localization compared to the case of a single chain
whereas repulsive ones {\it enhance} localization.
For the case with spin, two chains are always less localized than their
one chain counterpart. For each case we also compute physical quantities
such as localization length and conductivity.

Let us consider first two chains of spinless fermions
coupled by an interchain hopping $t_{\perp}$. Such model can also be
mapped to two spin chains coupled by an exchange X-Y
term, in the presence of a magnetic field.
For simplicity we will
just consider here a nearest neighbor interaction $V$.
More complicated interactions can be
considered without changing the main physical results. Details will be
given elsewhere \cite{orignac_2chain_long}. The disorder is modelled by
a random on-site potential $\epsilon_{i,p}$ uncorrelated from
site to site and from chain to chain. The Hamiltonian then reads
\begin{eqnarray}
\label{teve}
H & = & -t \sum_{i,p} c^{\dagger}_{i,p} c_{i+1,p} + h. c. +
V \sum_{i} n_{i,p} n_{i+1,p} \nonumber \\
 & & + t_{\perp} \sum_{i} c^{\dagger}_{i,1} c_{i,-1} + h. c. +
       \sum_{i,p} \epsilon_{i,p} n_{i,p}
\end{eqnarray}
where $p= -1,1$ is the chain index and $i$ is the site index.

It is convenient to rewrite the Hamiltonian in a standard boson
representation
\cite{emery_revue_1d,solyom_revue_1d,haldane_bosonisation}.
We therefore linearize the fermions dispersion relation
around $k_F$,
introduce right movers (R) and left movers (L) for each chain.
Then we take the continuum limit $c_{n,r,p}\rightarrow
\sqrt{\alpha}\psi_{r,p}(n\alpha)$ with $r=L,R $, $p=\pm1$
the chain index and $\alpha$ the lattice spacing. We rewrite
the hamiltonian in the bonding $\psi_{o}=\frac{\psi_{1}+
\psi_{-1}}{\sqrt{2}}$
and antibonding $\psi_{\pi}=\frac{\psi_{1}-\psi{-1}}{\sqrt{2}}$
bands base and
introduce the densities $\rho_{r,o,\pi}(x)=:\psi_{r,o,\pi}^{\dagger}(x)
\psi_{r,o,\pi}(x):$. We then define the canonically conjugate
fields $\phi_{\rho,\parallel}$ and $\Pi_{\rho,\parallel}$ via :
\begin{eqnarray} \label{fields}
\partial_{x}\phi_{\rho,\parallel} & = & -\frac{\pi}{\sqrt{2}}
(\rho_{L,o}+\rho_{R,o} \pm \rho_{L,\pi} \pm \rho_{R,\pi}) \\
\Pi_{\rho,\parallel} & = & \frac{1}{\sqrt{2}}(\rho_{R,o} \pm
\rho_{R,\pi}-\rho_{L,o} \mp\rho_{L,\pi})  \nonumber
\end{eqnarray}
and the field $\theta_{\rho,\parallel}(x)
=\int_{-\infty}^{x} \Pi_{\rho,\parallel}(x')dx'$.
In term of these fields the Hamiltonian becomes for the pure case
($\epsilon_{i,p}=0$) :
\widetext
\begin{eqnarray} \label{bosonise}
H & = & H_{\rho}+ H_{\parallel}, \qquad\qquad
H_{\rho} = \int \frac{dx}{2\pi}\left[ u_{\rho} K_{\rho}
(\pi \Pi_{\rho})^{2} +
\frac{u_{\rho}}{K_{\rho}}( \partial_{x} \phi_{\rho})^{2}\right]\\
H_{\parallel} &=& \int \frac{dx}{2 \pi} \left[u_{\parallel}
K_{\parallel} (\pi
\Pi_{\parallel})^{2}+\frac{u_{\parallel}}{K_{\parallel}}(\partial_{x}
\phi_{\parallel})^{2}\right] +\int dx t_{\perp} \frac{\sqrt{2}}{\pi}
\partial_{x}
\phi_{\parallel}
+ \int dx \left[\frac{2g_{\perp}}{(2 \pi \alpha)^{2}}
\cos(\sqrt{8}\phi_{\parallel})+ \frac{2g_{f}}{(2 \pi \alpha)^{2}}
\cos( \sqrt{8} \theta_{\parallel}) \right] \nonumber
\end{eqnarray}
\narrowtext
The expressions of the $K,u,g$ in terms of the
original parameters of the hamiltonian can easily be obtained
\cite{nersesyan_2ch,orignac_2chain_long}.
For the pure t-V model one has $K_\rho<1$
(resp. $K_\rho>1$) and $g_f < 0$ (resp. $g_f > 0$) for repulsive (resp.
attractive ) interactions and $K_{\parallel}=1$ for all $t,V$. By adding
interchain interactions,
one has access to the
cases $K_{\rho}>1$ and $g_{f}<0$ or $K_{\rho}<1$ and $g_{f}>0$.
The complete phase diagram in the pure case has been obtained in
\cite{nersesyan_2ch} by a mapping on a problem of one
chain of fermions with spin and spin-anisotropic interactions in a
magnetic field \cite{giamarchi_spin_flop}.
The $t_{\perp}$ term suppresses $\cos(\sqrt{8}\phi_{\parallel})$ so
that  $\theta_{\parallel}$ develops a gap and
acquires a non-zero expectation value determined by minimizing the
ground state energy.
The operators with divergent associated susceptibilities are then:
\widetext
\begin{eqnarray}
O_{CDW^{\pi}}& = &
\psi^{\dagger}_{R,o}(x)\psi_{L,\pi}(x)+\psi^{\dagger}_{R,\pi}
\psi_{L,o}(x)\sim
e^{\imath\sqrt{2}\phi_{\rho}}\cos(\sqrt{2}\theta_{\parallel})
\nonumber\\
O_{OAF} & = &
i(\psi^{\dagger}_{R,o}(x)\psi_{L,\pi}(x)-\psi^{\dagger}_{R,\pi}(x)
\psi_{L,o}(x))\sim
e^{\imath\sqrt{2}\phi_{\rho}}\sin(\sqrt{2}\theta_{\parallel})\nonumber\\
O_{S1} & = &
\psi_{L,o}(x)\psi_{R,\pi}+\psi_{L,\pi}\psi_{R,o} \sim
e^{\imath\sqrt{2}
\theta_{\rho}}\sin(\sqrt{2}\theta_{\parallel}) \nonumber\\
O_{S2} & = &
\psi_{L,o}\psi_{R,\pi}-\psi_{L,\pi}\psi_{R,o}
\sim e^{\imath\sqrt{2}\theta_{\rho}}\cos(\sqrt{2}\theta_{\parallel})
\nonumber
\end{eqnarray}
\narrowtext
These operators describe respectively out of phase charge density waves,
an orbital antiferromagnetic phase and chain symmetric ``s'' and
chain antisymmetric ``d'' type superconductivity.

For $g_f<0$ we have $\langle\theta_{\parallel}\rangle=0$ giving
an S2 phase for $K_{\rho}>1$
and the $CDW^{\pi}$ for $K_{\rho}<1$.
For $g_{f}>0$ we have
$\langle\theta_{\parallel}\rangle=\frac{\pi}{\sqrt{8}}$ giving
the S1 phase for $K_{\rho}>1$ and the OAF
phase for $K_{\rho}<1$.
In \cite{nersesyan_2ch} the bosonized
forms of $O_{S1}$ and $O_{S2}$ are exchanged due to the neglect of
anticommuting operators, so that the two superconducting phase
have been erroneously exchanged.

Now, we consider the effect of the disorder. Taking the
continuum limit for the on-site random potential, keeping only the
$2k_F$
terms in the bosonized expressions (as the forward scattering does not induce
localization \cite{abrikosov-rhyzkin}), and finally going to bonding and
antibonding bands, one finds that the coupling to disorder is represented
by two terms:
\begin{eqnarray}
H_{s} & = & \int\frac{dx}{\pi\alpha}\xi_{s}(x)
e^{\imath\sqrt{2}\phi_{\rho}}
\cos(\sqrt{2}\phi_{\parallel}) + h. c. \label{symmetric} \\
H_{a} & =& \int\frac{dx}{\pi\alpha}\xi_{a}(x)
e^{\imath\sqrt{2}\phi_{\rho}}
\cos(\sqrt{2}\theta_{\parallel}) + h. c. \label{antisymm}
\end{eqnarray}
where $\xi_{s,a}$ are two uncorrelated gaussian distributed random
potentials such
that $\overline{\xi_{n}(x)\xi_{n'}^{*}(x')}=D_{n}\delta_{n,n'}\delta(x-x')$
with $n,n'=s,a$.
In the original lattice problem, the role of $\xi_{s,a}$ would be played
respectively by $\epsilon_{n,1}\pm \epsilon_{n,-1}$.
We consider in the following a disorder weak enough not to destroy the
gaps opened by the interchain coupling in the pure system. This
corresponds to the limit $D \ll t_\perp$. The other limit where both the
interchain hopping and the disorder are small compared to the other
parameters by for arbitrary magnitude is only important in the vicinity
of the noninteracting point. It can be studied by similar methods and
will be discussed elsewhere \cite{orignac_2chain_long}.
In the weak disorder limit, $\phi_{\parallel}$ has huge quantum fluctuations,
and consequently $D_{s}$ will always be less relevant than $D_{a}$. We
can therefore concentrate on the latter and forget about the former.

First, we consider the case of $g_{f}<0$ (i.e. $V>0$ for the t-V model).
In that case, we can replace $\cos(\sqrt{2}\theta_{\parallel})$ by its
(non-zero) mean value.
Then the coupling to disorder (\ref{antisymm}) reduces to $C\int dx
\xi_{a}(x)e^{i\sqrt{2}\phi_{\rho}(x)}+ h. c. $ and the RG equations for that
problem have been derived and analysed in \cite{giamarchi_loc}. In
particular the disorder will grow under renormalization as
\begin{equation} \label{growth}
\frac{d D_a}{d l} = D_a (3 - K_\rho)
\end{equation}
where $l=\ln(\alpha)$ is the standard logarithmic scale associated with
cutoff renormalization.
(\ref{growth}) implies a localization-delocalization transition
\cite{giamarchi_loc} at $K_{\rho}=3$.
As a consequence, the d-wave superconducting phase is unstable
in the presence
of disorder except for extremely strong attractive interactions.
For a simple t-V model for which $K_\rho < 1$, the CDW ground state is
also unstable to disorder. The localized phase
is a Pinned Charge Density Wave phase, with a localization
length given by $L_{2 ch.}=(1/D)^{\frac{1}{3-K_{\rho}}}$. This is to be
compared to the localization length of a one dimensional spinless system
$L_{1 ch.}=(1/D)^{\frac{1}{3- 2 K_{\rho}}}$. For repulsive interactions the
effects of the interchain hopping
is therefore to decrease the localization length and
to make the two chains system more localized.
The conductivity above the pinning temperature $u/L_{2 ch.}$ can be
obtained by methods similar to \cite{giamarchi_loc} and varies as
$\sigma(T) \sim T^{2-K_\rho}$.

On the other hand if one considers $g_{f}>0$, i.e. attractive
interactions for a t-V model,
then $\langle \theta_{\parallel} \rangle = \frac{\pi}{\sqrt{8}}$ and in
a first approximation
$\langle \cos(\sqrt{2}\theta_{\parallel})\rangle=0$ so that there is
apparently no coupling at all to the disorder.
Obviously, such an approximation is too crude and
we must take into account the
fluctuations of $\theta_{\parallel} $ around its mean value.
Keeping only the relevant terms, and integrating
out the fluctuations of $\theta_{\parallel}$
around its mean value we obtain the following effective action
for $\phi_{\rho}$:
\widetext
\begin{eqnarray}
\label{action-finale}
S^{eff}_{\rho} &=& \int dx d\tau\left[ \frac{(\nabla\phi_{\rho})^{2}}{2\pi
K_{\rho}}+
\xi_{eff}(x)e^{\imath\sqrt{8}\phi_{\rho}(x,\tau)}+
\xi_{eff}^{*}(x)e^{-\imath
\sqrt{8}\phi_{\rho}(x,\tau)}\right]
\end{eqnarray}
\narrowtext
with $\overline{\xi_{eff}(x)\xi_{eff}^{*}(x')}=D_{eff}\delta(x-x')$
and $D_{eff}\sim D_{a}^2$.

The renormalization of the disorder will again be given by an equation
similar to (\ref{growth}), but with a coefficient $(3-4K_\rho)/2$ in front of
$D_{a}$.
The disorder is now relevant only for $K_{\rho}<3/4$, leading
to three different phases for $g_{f}>0$: a random orbital
antiferromagnet for $K_{\rho}<3/4$ (with a localization length
$L_{2 ch.}=(1/D)^{\frac{2}{3-4K_{\rho}}})$, an ordered orbital antiferromagnet
for
$3/4<K_{\rho}<1$ and a s-wave superconducting phase for $K_{\rho}>1$.
For the t-V model, $K_\rho > 1$, and the
``s''-wave superconducting phase is therefore {\em stable} with respect
to weak disorder, at variance to the single chain problem.
For the latter the delocalization only occured for extremely attractive
interactions i.e. $K_\rho > 3/2$. For the two chains problem
the localization-delocalization transition
arises in the immediate vicinity of the non-interacting point.
Contrarily to the case of repulsive interactions, interchain hopping
is now strongly reducing
the localization effects compared to the one dimensional case.
The determination of the critical properties at the boundary between the
repulsive (localized) regime
and the attractive (superconducting) one, requires to treat the case
where the gaps induced by the hopping and the disorder have arbitrary
relative strength \cite{orignac_2chain_long}. The conductivity now
behaves as $\sigma(T) \sim T^{2-4 K_\rho}$, and diverges as $T\to 0$
since the ground state is superconducting. In addition, since the
disorder is less relevant for attractive interactions than for
repulsive ones one can also expect the charge stiffness
\cite{kohn_stiffness}
for a disordered finite length two chains system to be larger
for the attractive case than for the
repulsive one, similarly to the one chain
\cite{bouzerar_spinless_currents,giamarchi_persistent_1d} system, but
with much more dramatic effects.

Let us consider now the problem with spins. Here again, we will for
simplicity only consider the case of a local Hubbard interaction. More
general interactions can be treated by the same method, and give rise
to a richer phase diagram \cite{schulz_2chains}.
The Hamiltonian is now
\begin{eqnarray}
H & = & -t\sum_{i,\sigma,p}c^{\dagger}_{i+1,\sigma,p}c_{i,\sigma,p} + h. c.
 - t_{\perp}\sum_{i,\sigma,p}c^{\dagger}_{i,\sigma,p}c_{i,\sigma,-p} \nonumber
\\
 & & +U\sum_{i,p} n_{i,\uparrow,p}n_{i,\downarrow,p}
 +\sum_{i,\sigma,p}\epsilon_{i,p}n_{i,\sigma,p}
\end{eqnarray}
The pure system can again be studied by using a boson representation.
One introduces similar fields than in (\ref{fields}) for each
spin degree of freedom, and make the symmetric (charge)
$\phi_\rho = \phi_\uparrow + \phi_\downarrow$ and antisymmetric (spin)
$\phi_\sigma = \phi_\uparrow - \phi_\downarrow$ linear combinations.
One ends with four bosonic fields instead of two for the spinless case.
The bosonized Hamiltonian is quite lengthy and will not be reproduced
here for reasons of space. It can be found in
\cite{schulz_2chains}, and we will use in the following the notations of
this paper. All physical quantities depends on a parameter
$K_{\rho+}$ of the symmetric charge mode,
analogous to the $K_\rho$ of the spinless problem.
For the purely repulsive case, $U >0$, only one
of the four bosonic fields that describe the low-energy physics of the
system ($\phi_{\rho +})$ is gapless \cite{schulz_2chains}.
The mean values of the 3 other fields are determined by
minimizing the energy
of the ground state giving $\langle \theta_{\rho -} \rangle =0$,
$\langle \phi_{\sigma +} \rangle =\frac{\pi}{2}$,
$\langle \phi_{\sigma -} \rangle =\frac{\pi}{2}$, leading to a
d-wave superconductive phase \cite{schulz_2chains}. For the attractive
case $U < 0$, $\phi_{\rho +}$ is again massless in the pure case. But
now, we have
$\langle \theta_{\rho -} \rangle = 0$, $\langle \phi_{\sigma +}
\rangle = 0$, $\langle \phi_{\sigma -} \rangle = 0$.
Here, the most divergent fluctuations are associated with the operator
$O_{SC^{s}} \sim e^{\imath\phi_{\rho +}}\cos(\phi_{\sigma +})
\cos(\phi_{\sigma -})$ which is the order
parameter for s-wave superconductivity.

The coupling to disorder arises again via two terms
\begin {eqnarray} \label {picdw}
H_{a} & = & \int\xi_{a}(x)O_{CDW^{\pi}}(x) + \xi_{a}^{*}(x)
O_{CDW^{\pi}}^{\dagger}(x) dx \\
\label{zerocdw}
H_{s} & = & \int\xi_{s}(x)O_{CDW^{o}}(x) +
\xi_{s}^{*}(x)O_{CDW^{o}}^{\dagger}(x) dx
\end{eqnarray}
Where $\overline{\xi_{n}(x)\xi_{n'}(x')^{*}}=D_{n}\delta_{n,n'}\delta(x-x')
(n,n'=a,s)$, the $\xi_{n}$ being random gaussian distributed potentials.
The operators $O$ represents charge density waves : $CDW^{o}$ is the in-phase
charge density wave, and $CDW^{\pi}$ is the out of phase one.

Assuming again that the disorder is weak enough not to destroy
the gaps, the $O$ operators have the simple form for repulsive
interactions
\begin{eqnarray}
\label{simpl_repuls}
O_{CDW^{o}} & \sim & e^{\imath \phi_{\rho +}} \sin(\phi_{\rho -}) \\
O_{CDW^{\pi}} & \sim & e^{\imath \phi_{\rho +}} \sin(\theta_{\sigma-})
\end{eqnarray}
These two operators have exponentially decaying
correlation functions and no direct coupling with disorder would exist
if one just took into account the mean values of the fields
$\phi_{\rho,-}$ and $\theta_{\sigma,-}$. As in the spinless case one
should integrate over fluctuations to get the effective coupling
\begin{equation}
\label{ougl}
S_{\rho +}^{\text{disorder}}=\int \xi_{\text{eff.}}(x)e^{\imath
2\phi_{\rho +}(x,\tau)} dx d\tau + h.c.
\end{equation}
One can also view (\ref{ougl})
as the coupling of the fermions with the $k_{Fo} \pm k_{F\pi}$ Fourier
component of the disordered potential.
The problem has in fact been reduced to a problem of spinless fermions.
The localization-delocalization would occur at $K_{\rho +}=3/2$ but purely
repulsive interaction imply $K<1$. The $d$-wave phase is therefore
unstable to abitrarily weak disorder.
The symmetric (\ref{zerocdw}) and the
antisymmetric (\ref{picdw}) part of the disorder contribute
equally to destroy the d-wave superconductivity, in contrast with the
spinless case where the antisymmetric part was the most relevant.
The localization length in that phase is $L_{2 ch.} \sim
(1/D)^{2/(3-2K_{\rho,+})}$, and therefore longer than the
corresponding one $L_{1 ch.}\sim (1/D)^{1/(2-K_{\rho,+})}$ of the one
dimensional spinning chain.
The two chains problem is less
localized than the corresponding one dimensional one even for
repulsive interactions, in contrast with the spinless case.
This is in qualitative agreement with what one expects in the absence of
interactions where the localization length is proportional to the number
of channels in the system.

For the attractive case, the $O$ operators take a different simplified
form, due to the different gaps in the system
\begin{eqnarray}
\label{simpl-o-attrac}
O_{CDW^{o}} & \sim &  e^{\imath\phi_{\rho +}}\cos(\phi_{\rho -}) \\
\label {simpl-pi-attrac}
O_{CDW^{\pi}} & \sim & e^{\imath\phi_{\rho +}} \sin(\theta_{\sigma -})
\sin(\phi_{\sigma +})
\end{eqnarray}
By substituting in
(\ref{picdw}) and (\ref{zerocdw}) and integrating over fluctuations
we end with an action of
the form (\ref{ougl}). This time, $K_{\rho,+}>1$ , so that we can attain
the localization-delocalization transition at $K=3/2$. The
delocalization transition arises for much weaker attraction than in the one
dimensional case \cite{giamarchi_loc}
$K_{\rho}=3$. For the two chains problem the
critical value of $K$ can be attained for a Hubbard model
\cite{kawakami_bethe_U<0,giamarchi_persistent_1d}
whereas the
one dimensional Hubbard model is always localized even for very negative
$U$ \cite{giamarchi_persistent_1d}.
In addition the localization length is increased : $L_{2 ch.}=
(\frac{1}{D})^{\frac{2}{3-2K_{\rho +}}}$ whereas in the one chain case
$L_{1 ch.}=(\frac{1}{D})^{\frac{1}{3-K_{\rho}}}$.
At
the opposite to what happens for the one dimensional case where the
attractive localization length was smaller than the repulsive one
\cite{suzumura_scha,giamarchi_loc}, here the two lengths are the same, up
to prefactors. Therefore, the enhancement of charge stiffness by repulsive
interactions found in the one chain case \cite{giamarchi_persistent_1d}
ought to be absent for 2 chains, or at least strongly reduced.
Such an issue would need a more detailed study. The conductivity
behaves as $\sigma(T)\sim T^{2-2K_{\rho+}}$.

Clearly, these effects are due to the existence of a spin gap and
to the freezing of interchain charge excitations\cite{schulz_2chains} . As a
consequence, it would
be worth studying the localization effects in a three chains model
(where there should be no spin-gap) to see if the delocalization
effect of attractive
interaction does persist or if we fall back to the one chain case.

We are grateful to H.J. Schulz for many useful discussions.

{\it Note added:} After completion of this work, we learned about the
work of Kawakami and Fujimoto \cite{kawakami_dopeds=1}. These authors
considered the
related, albeit different problem of disordered coupled
Hubbard chains with a ferromagnetic Hund's exchange and no hopping.
They also find reduction of the localization effects in this system.

%\bibliographystyle{prsty}
%\bibliography{revues,total,2chain}

%\end{multicols}
\end{document}